\documentclass{aa}
\input{psfig.sty}
\usepackage{lscape}
\usepackage{graphicx,times}
\usepackage{float}
\usepackage{longtable}

\begin{document}
\authorrunning{K. C. Steenbrugge et al.}
\titlerunning{Spectroscopy of NGC~5548 with XMM-NEWTON}
\title{XMM-NEWTON High Resolution Spectroscopy of NGC~5548}
\subtitle{}

\author{K.C. Steenbrugge\inst{1}, J.S. Kaastra\inst{1}, C. P. de Vries\inst{1}, R. Edelson\inst{2}}
\offprints{K.C. Steenbrugge}
\mail{K.C.Steenbrugge@sron.nl}
\institute{ SRON National Institute for Space Research, 3584 CA Utrecht, The Netherlands
\and Department of Physics and Astronomy; University of California, Los Angeles; Los Angeles, CA 90095-1562, USA}
\date{Received 21 August 2002 / Accepted 14 February 2003}

\abstract{We analyze a 137 ks exposure X-ray spectrum of the Seyfert 1 galaxy NGC~5548 obtained with the XMM-{\it Newton} Reflection Grating Spectrometer. Due to the long exposure time, the spectrum is of higher statistical quality than the previous observations of this AGN. Therefore, we detect for the first time in NGC~5548 inner-shell transitions from \ion{O}{iii} to \ion{O}{vi} ions, and the Unresolved Transition Array of M-shell iron. The warm absorber found from this X-ray observation spans three orders of magnitude in ionization parameter. We detect \ion{O}{iii}, which is as lowly ionized as the warm absorber detected in the UV band, to \ion{Fe}{xxiv}. For \ion{O}{vi} the column density determined from our X-ray data is an order of magnitude larger than the column density measured in previous UV observations. We conclude that there is substantially more low ionized material than previously deduced from UV observations. However, only a few percent of the warm absorber detected in the X-rays is lowly ionized. A 99.9\% significant increase in the derived absorbing column density with higher ionization states is observed. The outflow velocities determined from the X-ray absorption lines are consistent with those deduced from the UV lines, evidence, together with the detection of \ion{O}{vi}, that the X-ray and UV warm absorber are different manifestations of the same phenomenon. From a simple mass conservation argument, we indicate that our data set is consistent with an outflow with small opening angle formed due to instabilities in the accretion disk. Possible due to uncertainties in the radiative transport mechanism, an apparent deviant iron to oxygen abundance is detected. No strong relativistically broadened emission lines of \ion{O}{viii}, \ion{N}{vii} and \ion{C}{vi} were detected.
}

\maketitle

\section{Introduction}
\label{sec:intro}
The immediate environment of active galactic nuclei (AGN) is poorly understood. The black hole is fed from an accretion disk, which becomes visible through high energy radiation. Possibly powered by radiation pressure, an outflowing wind is formed from the accretion disk. In some AGN, the wind is visible through its continuum and line absorption of the radiation from the nucleus. These absorption lines were already detected in the UV part of the spectrum (Mathur, Elvis \& Wilkes \cite{mathur}), and have now been confirmed in the X-ray spectra (Kaastra et al. \cite{kaastra00}). 
\par 
NGC~5548, a Seyfert 1 galaxy, has been well studied in the X-rays due to its relative X-ray brightness, and the fact that it is a nearby active  galactic nucleus with a redshift of z=0.01676 (Crenshaw \& Kraemer \cite{crenshaw}). This Seyfert galaxy was studied in detail with ROSAT (Done et al. \cite{done}). They detected an \ion{O}{vii} and \ion{O}{viii} K-shell absorption edge, and concluded that there was a warm absorber present. Therefore, NGC~5548 is an ideal Seyfert 1 galaxy to study with the high resolution spectroscopy instruments on-board of XMM-{\it Newton} and {\it Chandra}. Table~\ref{tab:obs} list the observations made with these instruments. 
In this paper we discuss a 137 ks exposure time Reflection Grating Spectrometer (RGS) spectrum from the guest observation program of XMM-{\it Newton}.
 
\section{Observations and data reduction}
Here we present data taken with the RGS instrument (den Herder et al. \cite{denherder}). The EPIC data (the continuum spectrum, and the Fe K$\alpha$ line) are presented separately (Pounds et al. \cite{pounds}). The time variability of the combined XMM-{\it Newton} and RXTE data sets will be presented by Markowitz et al. (\cite{mark}).

\begin{table}
\caption{High resolution X-ray observations of NGC~5548. LETGS: Low Energy Transmission Grating in combination with the High Resolution Camera; HETGS: High Energy Transmission Grating in combination with the ACIS camera both onboard {\it Chandra}; RGS: Reflection Grating Spectrometer onboard XMM-{\it Newton}. The exposure time is given in in ks.}
\label{tab:obs}
\begin{tabular} {|llll|}
\hline
Date         & Instr. & Expos. & References-comments     \\\hline
1999 Dec. 11 & LETGS  & 86.4   & Kaastra et al. 2000, 2002 \\
2000 Feb. 5  & HETGS  & 83     & Yaqoob et al. 2001, \\
             &        &        & Kaastra et al. 2002 \\
2000 Dec. 24 & RGS    & 28     & noisy spectrum \\
             &        &        & low flux state \\
2001 Jul. 09--12 & RGS    & 137    & present work     \\\hline
\end{tabular}
\end{table}
\par
Shortly after launch two of the eighteen CCD chains failed. For RGS 1 this causes a data gap in the $10-14$ \AA~range (where most Fe-L and Ne absorption lines are), for RGS 2 the gap is between $20-24$ \AA~(where the \ion{O}{vii} triplet is). 
\par
The data were reduced using the developer's version of the XMM-{\it Newton} data analysis package SAS version 5.3, which comprises a correct calibration of the effective area and the instrumental O-edge. The latter is important as the 1s-2p absorption lines of \ion{O}{ii} and \ion{O}{iii} are cosmologically redshifted toward the instrumental \ion{O}{i} edge. 
\par 
In this paper we analyze the first and second order spectra obtained from both RGS 1 and 2. Although the second order spectrum has a lower count rate and only a wavelength coverage from 5 to 19 \AA, it does have a higher wavelength resolution than the first order.
Throughout the paper we have fitted RGS 1 and 2 simultaneously, but for clarity of representation we have added the RGS 1 and RGS 2 spectra in the figures.
\par
\begin{figure}
\resizebox{\hsize}{!}{\includegraphics[angle=-90]{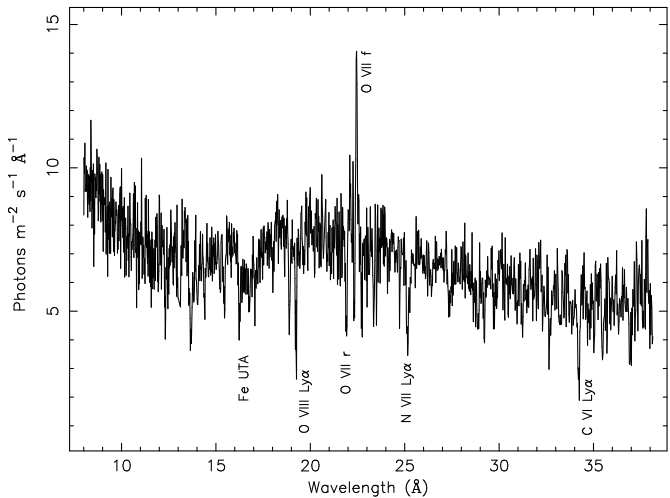}}
\caption{The fluxed spectrum of NGC~5548, RGS 1 and 2 are added together and no error bars are plotted for clarity. The typical 1 $\sigma$ error on the data is 1 photon m$^{-2}$ s$^{-1}$ \AA$^{-1}$. Some of the strongest features are identified.}
\label{fig:flux}
\end{figure}

All the spectral analysis was done with the SPEX package (Kaastra et al. \cite{kaastrasp}). Fig.~\ref{fig:flux} shows the fluxed spectrum of NGC~5548, indicating the Unresolved Transition Array (UTA) of inner-shell iron absorption lines and some other prominent features. 

\subsection{Wavelength scale accuracy}
In general the RGS wavelength scale can be reconstructed to an accuracy of 8 m\AA~(den Herder et al. \cite{denherder}). However, in our case larger uncertainties exist, due to errors in the recorded satellite attitude and right ascension in the data files available. From the analysis of the first order data we find that the \ion{O}{vii} forbidden line is blueshifted by 285 km s$^{-1}$ ($-$0.021 \AA). The \ion{O}{vii} forbidden line is not blueshifted in the earlier LETGS data of NGC~5548 (Kaastra et al. \cite{kaastra02}). No blueshift for the forbidden line can be determined in second order. However, comparing first and second order results for the warm absorber, we find a smaller outflow velocity in the second order data set. The difference in wavelength between first and second order, for the strongest line in the second order, \ion{O}{viii} Ly$\beta$ (both orders were fitted with the {\it xabs} model, see Sect. 3) is $-$0.019 $\pm$ 0.005 \AA. Finally, in the first order data set there is a weak Galactic \ion{O}{vii} recombination line, which has a rest wavelength of 21.6 \AA. This absorption line also shows an apparent velocity shift of $-$230 km s$^{-1}$ ($-$0.017 \AA), although no blueshift should be expected. To correct for this wavelength calibration problem we applied a constant wavelength shift of $-$0.017 \AA~to the model for the first order spectra. If not corrected, this wavelength shift would result in a velocity error between 130 km s$^{-1}$ for the longest wavelengths to 620 km s$^{-1}$ for the shortest wavelengths. The residual error in the wavelength scale should be within the 8 m\AA~uncertainty of the calibration. All velocities quoted in this paper are corrected for this wavelength shift.

\section{Modeling the warm absorber}

In NGC~5548 there is an intricate pattern of densely spaced absorption lines as well as the possible presence of weak relativistic emission lines (Kaastra et al. \cite{kaastra02}). Blending, mostly of iron lines, is important in many wavelength ranges, making equivalent width (EW) estimations rather complicated. On the other hand, in the X-ray band many of the observed ions have more than one detected absorption line, significantly improving the column density and velocity outflow measurements. This is also important in order to disentangle possible partial covering factors and column densities of the ions. The possibility of relativistic emission lines complicates the analysis of the continuum spectral shape, which is already complicated to derive due to the presence of many edges and absorption lines. We consider here two approaches for fitting the warm absorber. In the first approach all absorption lines and edges of each individual ion are calculated, using a separate value for the column density for each ion. The transmission of all ions is then combined to calculate the total transmission of the warm absorber. Outflow velocity and velocity broadening are coupled for the different ions. This is the {\it slab} model in SPEX (Kaastra et al. \cite{kaastrasp}). 
\par
In the second approach the column densities of the ions are coupled through a photoionization model. In this case only the total column density, ionization parameter, elemental abundances and the outflow and broadening are free parameters. This is the {\it xabs} model. For a full description of both methods see Kaastra et al. (\cite{kaastrasp}). 
\par
The results obtained with the {\it xabs} model are more dependent on theory, because the ionization parameter determines the column density ratios of the ions. 
An advantage of the {\it xabs} model is that different ionization components, each with their own velocity structure can be combined. This is harder to accomplish with {\it slab} components, in particular if the velocity components are barely separable with the current spectral resolution. This is the case in NGC~5548, and would lead to strongly correlated parameters in the spectral fitting procedure. One further difference between the models is that ions which have a rather small cross-section are automatically included in {\it xabs}, but since these are more difficult to discern individually, they are not usually included in the {\it slab} modeling. 
\section{Data analysis}
\subsection{The continuum}
\begin{table}
\caption{The parameter values quoted are those obtained in the {\it xabs} model fitting and differ from the best fit continuum without absorption. The continuum model was used as a basis for adding the warm absorber, either with a {\it slab} component (see Table~\ref{tab:slabfit}) or with {\it xabs} components (see Table~\ref{tab:xabsfit}). The quoted ${\chi_\nu}^2$ is for the best fit continuum parameters without absorption, and thus not the parameters quoted in the table, which we derived including absorption.}
\label{tab:contfit}
\begin{tabular} {|l|l|c|}
\multicolumn{3}{c}{${\chi_\nu}^2$ equals 1.96 for 2441 degrees of freedom}\\\hline
pl:  & norm$^{a}$  & (1.57 $\pm$ 0.03) $\times$ 10$^{52}$ ph s$^{-1}$ keV$^{-1}$ \\
     & $\Gamma$    &  1.77 $\pm$ 0.02 \\
mbb: & norm$^{b}$  & (6.5 $\pm$ 0.7) $\times$ 10$^{32}$ m$^{1/2}$ \\
     & $T$         & (97 $\pm$ 6) eV \\\hline
\end{tabular}\\
$^{a}$ at 1 keV.\\
$^{b}$ norm=emitting area times square root of electron density.\\
\end{table}
The continuum model consists of a power-law spectrum with Galactic absorption, and a modified black body (see Kaastra \& Barr \cite{kaastrab} for more details). A second modified black body with a high temperature is required to describe the EPIC data, but is not detected in the limited RGS band (Pounds et al. \cite{pounds}). This indicates that RGS data alone cannot constrain the continuum accurately. Table~\ref{tab:contfit} gives the continuum parameters used for the analysis of the RGS data.
\par
The temperature ($T$) of the modified black body was allowed to vary as well as the normalizations of the power-law and the modified black body. However, the photon index ($\Gamma$), which is very sensitive to small changes in the model, as well as the Galactic \ion{H}{i} column density (1.65 $\times$ 10$^{24}$ m$^{-2}$, Nandra et al. \cite{nandra}) were kept fixed throughout the fitting process. The obtained parameters are within calibration uncertainties consistent with those obtained from simultaneous EPIC and BeppoSAX observations.

\subsection{Spectral fit using the {\it slab} model}
\begin{table*}
\caption{Best fit values for the logarithms of the column densities in m$^{-2}$ and the outflow velocity, using the {\it slab} component. The logarithm of the ionization parameter for which the ion has the largest column density as determined from XSTAR is given in column 4.}
\label{tab:slabfit}
\begin{center}
\begin{tabular}{|lcrr|lcrr|}
\hline
               & log N$_{\rm H}$ & $v$              & log $\xi$ &                & log N$_{\rm H}$ & $v$              & log $\xi$      \\
ion            & (m$^{-2}$)      & (km s$^{-1}$)    & (10$^{-9}$ W m)& ion            & (m$^{-2}$)      & (km s$^{-1}$)    & (10$^{-9}$ W m) \\\hline
\ion{C}{iv}$^{a}$& 21.4 $\pm$ 0.1  & $-$      & $-$ 0.6   & \ion{S}{xiv}   & 20.1 $\pm$ 0.3  & +90 $\pm$ 280    & 2.4            \\
\ion{C}{v}     & 20.4 $\pm$ 0.2  & $-$490 $\pm$ 200 & 0.2       & \ion{Ar}{xiii} & 19.4 $\pm$ 0.4  & $-$630 $\pm$ 210 & 1.8            \\
\ion{C}{vi}    & 21.1 $\pm$ 0.1  & $-$510 $\pm$ 70  & 1.2       & \ion{Ar}{xiv}  & $<$ 19.4        & $-$              & 2.1            \\
\ion{N}{v}$^{a}$& 20.6 $\pm$ 0.3  & $-$       & 0.0       & \ion{Ar}{xv}   & 20.1 $\pm$ 0.3  & $-$470 $\pm$ 400 & 2.3            \\
\ion{N}{vi}    & 20.1 $\pm$ 0.1  & $-$470 $\pm$ 130 & 0.7       & \ion{Fe}{i}    & $<$ 19.8        & $-$              & $< -$ 4.0      \\
\ion{N}{vii}   & 20.5 $\pm$ 0.1  & $-$390 $\pm$ 150 & 1.5       & \ion{Fe}{ii}   & 19.9 $\pm$ 0.3  & $-$              & $-$ 3.0        \\
\ion{O}{iii}   & 20.0 $\pm$ 0.6  & $-$              & $-$ 1.8   & \ion{Fe}{v}    & $<$ 19.7        & $-$              & $-$ 1.3        \\
\ion{O}{iv}    & $<$ 20.2        & $-$              & $-$ 0.7   & \ion{Fe}{vi}   & 19.8 $\pm$ 0.4  & $-$              & $-$ 0.9        \\
\ion{O}{v}$^{b}$ & 20.7 $\pm$ 0.3  & +430           & 0.0       & \ion{Fe}{vii}  & $<$ 19.7        & $-$              & $-$ 0.3        \\
\ion{O}{vi}    & 20.5 $\pm$ 0.1  & $-$250 $\pm$ 160 & 0.5       & \ion{Fe}{viii} & 20.1 $\pm$ 0.1  & $-$340 $\pm$ 450 & 0.1            \\
\ion{O}{vii}   & 21.57 $\pm$ 0.05 & $-$450 $\pm$ 90 & 1.2       & \ion{Fe}{ix}   & 20.1 $\pm$ 0.1  & +130 $\pm$ 220   & 0.4            \\
\ion{O}{viii}  & 22.24 $\pm$ 0.03 & $-$310 $\pm$ 70 & 1.8       & \ion{Fe}{x}    & 19.9 $\pm$ 0.1  & $-$670 $\pm$ 430 & 0.6            \\
\ion{Ne}{ix}   & 21.82 $\pm$ 0.06 & $-$30 $\pm$ 200 & 1.7       & \ion{Fe}{xi}   & 20.1 $\pm$ 0.1  & $-$430 $\pm$ 400 & 0.8            \\
\ion{Ne}{x}    & 21.5 $\pm$ 0.2  & +10 $\pm$ 230    & 2.3       & \ion{Fe}{xii}  & $<$ 19.7        & $-$              & 1.0            \\
\ion{Na}{x}    & 20.5 $\pm$ 0.4  & $-$              & 1.9       & \ion{Fe}{xiii} & 19.4 $\pm$ 0.6  & $-$              & 1.2            \\
\ion{Na}{xi}   & 20.6 $\pm$ 0.6  & $-$              & 2.4       & \ion{Fe}{xiv}  & $<$ 19.5        & $-$              & 1.4            \\
\ion{Mg}{x}    & 21.4 $\pm$ 0.1  & $-$560 $\pm$ 250 & 1.5       & \ion{Fe}{xv}   & $<$ 19.7        & $-$              & 1.6            \\
\ion{Mg}{xi}   & 20.5 $\pm$ 0.5  & $-$              & 2.1       & \ion{Fe}{xvi}  & 19.7 $\pm$ 0.3  & $-$500 $\pm$ 450 & 1.6            \\
\ion{Mg}{xii}  & 20.8 $\pm$ 0.5  & $-$970 $\pm$ 990 & 2.6       & \ion{Fe}{xvii} & 20.1 $\pm$ 0.1  & $-$480 $\pm$ 170 & 2.1            \\
\ion{Si}{ix}   & 21.3 $\pm$ 0.1  & $-$              & 0.9       & \ion{Fe}{xviii}& 20.1 $\pm$ 0.1  & $-$420 $\pm$ 240 & 2.3            \\
\ion{Si}{x}    & $<$ 20.4        & $-$              & 1.3       & \ion{Fe}{xix}  & 20.49 $\pm$ 0.09 & $-$350 $\pm$ 230& 2.5            \\
\ion{Si}{xi}   & $<$ 20.1        & $-$              & 1.7       & \ion{Fe}{xx}   & 20.4 $\pm$ 0.2  & $-$960 $\pm$ 710 & 2.8            \\
\ion{Si}{xii}  & $<$ 20.5        & $-$400 $\pm$ 380 & 2.1       & \ion{Fe}{xxi}  & 20.5 $\pm$ 0.2  & $-$65 $\pm$ 550  & 3.0            \\
\ion{S}{x}     & 20.3 $\pm$ 0.2  & $-$820 $\pm$ 420 & 1.1       & \ion{Fe}{xxii} & 20.6 $\pm$ 0.2  & $-$              & 3.2            \\
\ion{S}{xii}   & 20.3 $\pm$ 0.1  & $-$410 $\pm$ 80  & 1.7       & \ion{Fe}{xxiii}& 20.4 $\pm$ 0.4  & $-$              & 3.3            \\
\ion{S}{xiii}  & 20.1 $\pm$ 0.2  & $-$640 $\pm$ 200 & 2.1       & \ion{Fe}{xxiv} & 20.7 $\pm$ 0.5  & $-$              & 3.5            \\\hline
\end{tabular}\\
\end{center}
$^{a}$ See text for details.\\
$^{b}$ Possible error in the wavelength.\\
\end{table*} 
\par
As a first approach to the analysis of the warm absorber, we modeled it by adding a {\it slab} component to the continuum model. We refitted the data, leaving the same parameters free as above for the continuum, as well as the overall outflow velocity, and all the relevant ion column densities. Finally, after obtaining this overall fit we also let the photon index vary, and now obtain a reasonable result. The results of this fit were then used to fine-tune the column density and outflow velocity of each ion which has a detectable absorption line or edge in the RGS wavelength range. To do so we decoupled subsequently for each ion its outflow velocity from the outflow velocity of all the other ions in the model. This method was followed to minimize the effects from blending. The significance of the different ionic columns can be derived from the best fit column densities and their error bars. The result of this fitting is summarized in Table~\ref{tab:slabfit}. This Table also lists the ionization state for the maximum column density of the particular ion. Note however, that each ion exists for a range of ionization states, and thus the quoted ionization is only an indication. Adding the {\it slab} component significantly improves the ${\chi_\nu}^2$ from 1.96 to 1.26 (with 2404 degrees of freedom). 
\par
For the ions with only an upper limit for their column densities, absorption is only detected through continuum absorption (\ion{Ar}{xiv}, \ion{Fe}{v} and \ion{Fe}{vii}) or a single/few weak lines (\ion{O}{iv}, \ion{Si}{x}, \ion{Si}{xi}, \ion{Fe}{i}, \ion{Fe}{xii} and \ion{Fe}{xiv}). As a result the column density as well as the velocity of these ions are poorly determined. The other ions with poorly determined velocity shifts, but with solid column densities, are those for which the column density is determined from an edge (\ion{C}{iv}, \ion{N}{v} and \ion{Si}{ix}), for which the absorption lines are strongly blended (\ion{Fe}{xii}), or because only a few weak absorption lines are detected.
\par
We find a very well constrained column density for \ion{C}{iv} and \ion{N}{v}, although both these ions have in the RGS band only an absorption edge, at 36 \AA~and 25 \AA~(rest wavelength) respectively. As the depth of the edge correlates with the assumed underlying continuum, these column densities are rather more uncertain than the fit results indicate. To ascertain how these column densities change the overall continuum shape we refitted the data, but now forcing a column density for \ion{C}{iv} and \ion{N}{v} of zero. For the new fit, which gives a similarly good ${\chi_\nu}^2$, the normalization of the modified black body increased by an order of magnitude, while the temperature of the same component decreased by an order of magnitude. As a result of this change in the continuum model, the column densities found for \ion{O}{iii} and \ion{Ne}{ix} also changed significantly. In this respect the {\it xabs} modeling is more reliable, as the  column densities determined for ions that are only detected through an edge, are based upon the determined ionization state.
\par
\begin{figure}
\resizebox{\hsize}{!}{\includegraphics[angle=-90]{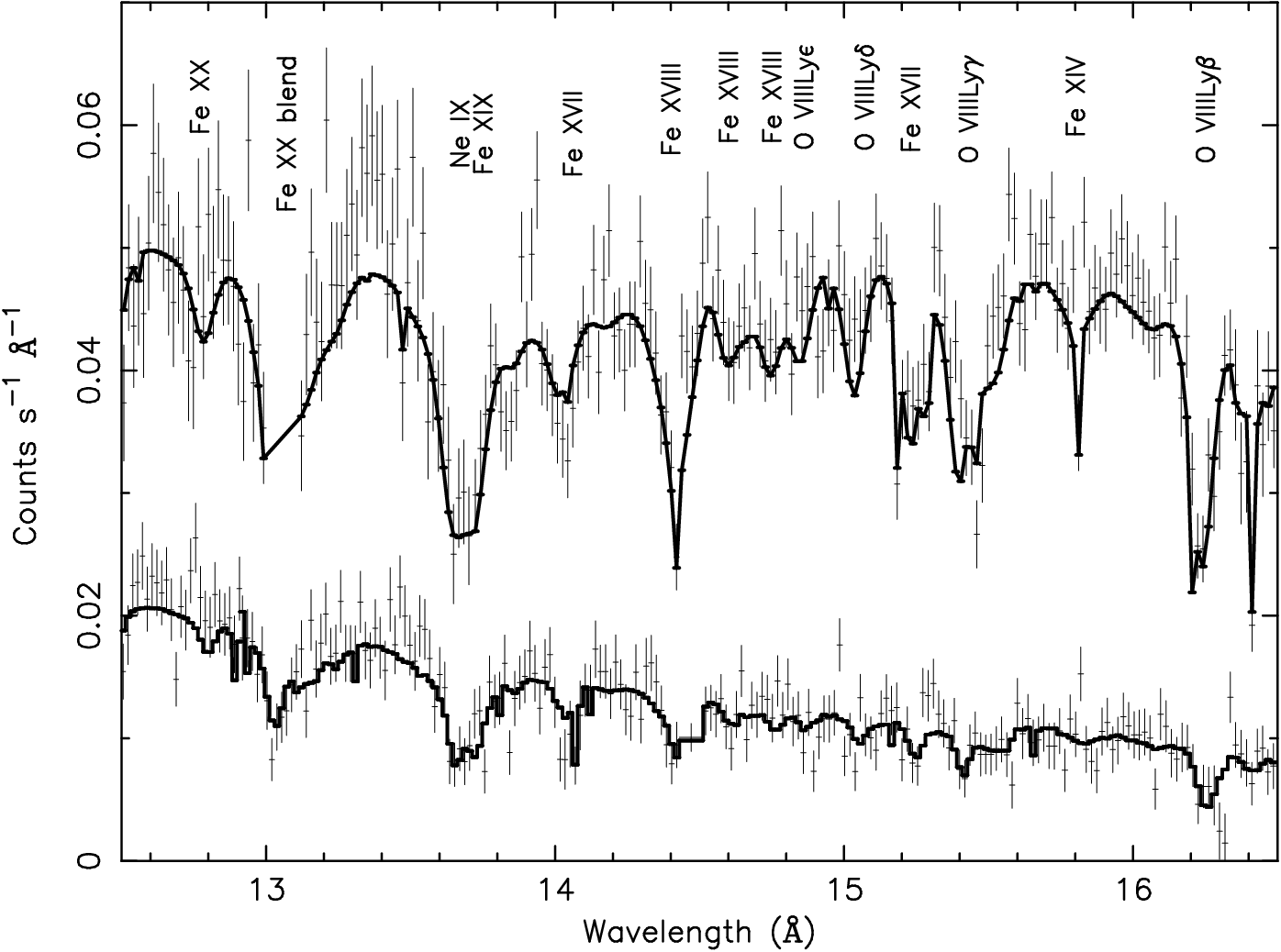}}
\caption{Best {\it xabs} fit to the first (top) and second order (bottom) data. First order data are rebinned by a factor of 2, second order by a factor of 3. For the first order data a CCD gap related feature can be seen at 13 \AA~and a dead pixel is seen short-ward of 16.5 \AA. Between 12.5 and 14 \AA~only RGS 2 gives results, hence the larger error bars. For second order data a dead pixel is seen at 14.05 \AA~and a CCD gap related feature can be seen at 14.45 \AA.}
\label{tab:o1fit}
\end{figure}

The same method for modeling the warm absorber was used for fitting the second order data. The second order data for RGS 1 and 2 were not fitted together with the first order data, as the systematic shift in  the first order spectra is corrected for in the applied model and not in the data itself. However, both orders are in excellent agreement, see Fig.~\ref{tab:o1fit}. 
\par
The average outflow velocity found, including only those ions with well determined velocities in both orders, is $-$350 $\pm$ 250 km s$^{-1}$ for the first order data set, in comparison with $-$420 $\pm$ 340 km s$^{-1}$ for the second order data set. For further analysis we focus on the first order data, because the better statistics outweigh the better spectral resolution of the second order data.

\subsection{Spectral fit using the {\it xabs} model}
\begin{table}
\caption{The best fit results for the first order data {\it xabs} fit. Further details are given in the text.}
\label{tab:xabsfit}
\begin{tabular} {|l|ccc|}\hline
Component          & A        & B       & C                       \\\hline
log $\xi$ (10$^{-9}$ W m)& 2.69$\pm$0.04   & 1.98$\pm$0.06  & 0.40$\pm$0.03 \\
log $N$$_{\rm H}$ (m$^{-2}$) & 25.68$\pm$0.05 & 25.52$\pm$0.06 & 24.15$\pm$0.01 \\
outflow $v$ (km s$^{-1}$) & $-$311$\pm$60 & $-$440$\pm$100 & $-$290$\pm$70 \\\hline
abundances$^{a}$:  &            &             &                            \\
C    & $<$4.9      & 0.6$\pm$0.2    & 1.8$\pm$0.6              \\
N    & 2.2$\pm$1.0 & $<$0.3         & 1.7$\pm$0.5              \\
Ne   & 2.5$\pm$1.4 & 0.1$\pm$0.7    & b                          \\
Na   & $<$8.3      & 3.9$\pm$3.9    & $<$350                     \\
Mg   & 1.0$\pm$0.8 & $<$0.6         & $<$5.8                    \\
Si   & 2.2$\pm$2.2 & $<$0.2         & $<$3.9                    \\
S    & 0.6$\pm$0.7 & 0.5$\pm$0.2    & b                          \\
Ar   & 2.2$\pm$2.2 & 0.4$\pm$0.5    & $<$5.4                    \\
Fe   & 0.70$\pm$0.06 & 0.11$\pm$0.03   & 7.4$\pm$0.6              \\\hline
\end{tabular}\\
$^{a}$ The abundances are relative to O, which is kept at solar value.\\
$^{b}$ Frozen to solar value, see text for explanation.\\
\end{table}
Using the same continuum model as in the previous section, we replaced the {\it slab} model by the {\it xabs} model for the warm absorber. First we fitted only the ionization parameter ($\xi$), the hydrogen column density ($N$$_{\rm H}$) and the outflow velocity ($v$), in addition to the continuum parameters that we leave free throughout the fitting procedure.  The abundances were kept to the solar values (Anders \& Grevesse \cite{anders}). Later, the abundances were left as free parameters. However, as hydrogen produces no X-ray lines, we normalized the abundance relative to oxygen, which was frozen to solar abundance. Finally we added two more {\it xabs} components to the model, both with lower ionization parameters, significantly improving the spectral fit. 
\par
Adding the first {\it xabs} component improved ${\chi_\nu}^2$ from 1.97 to 1.66. The addition of the second {\it xabs} component, which was the one with the lowest ionization parameter, further improved ${\chi_\nu}^2$ to a value of 1.46. However, for this {\it xabs} component we found two abundances which have unphysical values. Namely, Ne and S are overabundant by a factor of 43 and 106, respectively. From detailed fitting, it is clear that both these abundances are mainly determined from the absorption edges in the spectrum, and are thus rather uncertain. Fitting the model again, but fixing the abundance of Ne and S in the lowest ionization component to solar, we find a similar good fit, namely with ${\chi_\nu}^2$ of 1.48. The addition of the last {\it xabs} component gives a final ${\chi_\nu}^2$ = 1.22. Table~\ref{tab:xabsfit} lists the best fit parameters, with the {\it xabs} components labeled to be consistent with the components found by Kaastra et al. (\cite{kaastra02}). The significance, according to F-tests, is at least 100, 99.8 and 99.9 percent for components A, B and C, respectively. Only in the last step we left the photon index free, but this did not improve the ${\chi_\nu}^2$. The continuum parameters for this model correspond within 1 $\sigma$ with the continuum parameters as found from the {\it slab} fit. 

\subsection{Emission and absorption by oxygen}
\begin{figure}
\resizebox{\hsize}{!}{\includegraphics[angle=-90]{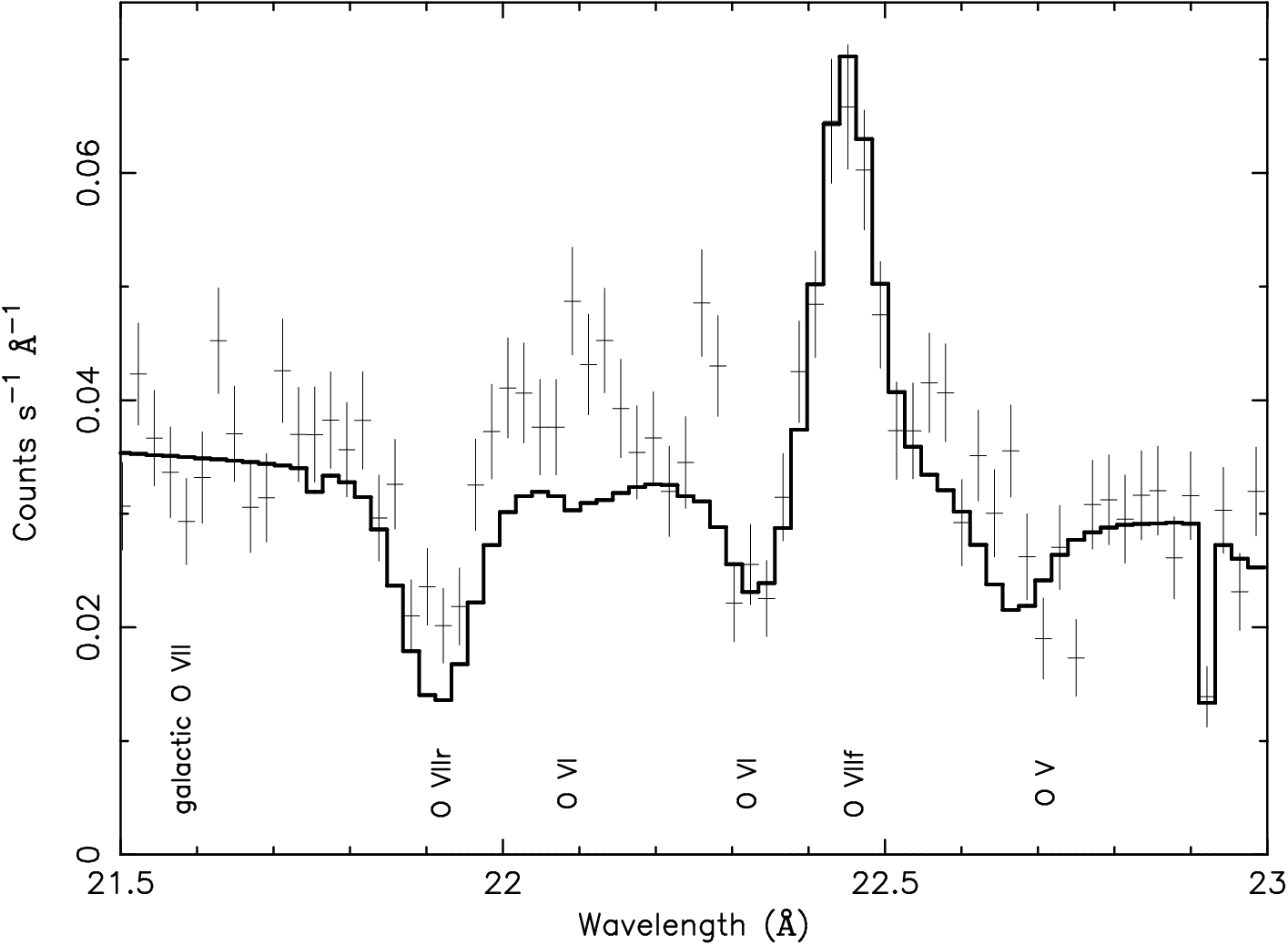}}
\caption{Detail of RGS 1 showing the \ion{O}{vii} triplet as well as absorption lines for \ion{O}{vi} and \ion{O}{v}. The {\it xabs} model was used for fitting the warm absorber. The Galactic \ion{O}{vii} absorption line is indicated.}
\label{fig:oreg}
\end{figure}

Fig.~\ref{fig:oreg} shows the spectrum around the \ion{O}{vii} triplet. In Table~\ref{tab:ofit} we list the parameters of the \ion{O}{vii} forbidden line in emission. For the \ion{O}{vii} forbidden line Table~\ref{tab:como7} gives the fluxes, outflow velocity as well as an upper limit to the width of the line, as measured from the HETGS and LETGS spectra (data taken from Kaastra et al. \cite{kaastra02}) and our data set. Note that this line has the same measured flux over a period of 1.5 years, despite large variations in the continuum. In all three observations the line is unresolved. The above indicate that the \ion{O}{vii} forbidden line is formed at least 0.46 pc from the ionizing source.
\begin{table}
\caption{Values determined for the forbidden line of \ion{O}{vii}.}
\label{tab:ofit}
\begin{tabular} {|r|l|c|}
\hline
\ion{O}{vii}f: & EW & ($-$170 $\pm$ 22) m\AA \\
               & flux & (0.9 $\pm$ 0.1) ph m$^{-2}$ s$^{-1}$ \\
               & FWHM & $<$ 36 m\AA \\
               &      & $<$ 490 km s$^{-1}$ \\
               & $\lambda$$^{a}$ & (22.096 $\pm$ 0.005) \AA \\
               & $v$$^{b}$ & ($-$110 $\pm$ 80) km s$^{-1}$ \\\hline
\end{tabular}\\
$^{a}$ Wavelengths are given in the rest frame of NGC~5548.\\
$^{b}$ Velocity shift from comparing rest and observed wavelengths.\\ 
\end{table}
\par
Interestingly, the inter-combination line is not present in the data. An excess is observed at 21.74 \AA~rest wavelength, shifted by 0.06 \AA~from the rest wavelength of the inter-combination line. The relative wavelength accuracy of RGS is 2 m\AA. Assuming that the inter-combination and forbidden lines are formed at the same distance from the ionizing source and under the same conditions, a blueshift of 620 km s$^{-1}$ is inconsistent with the lack of blueshift for the forbidden line. Because the forbidden line is strong, models predict that even tenuous gasses should produce a weak but detectable inter-combination line. One possible explanation is that the inter-combination line blends with an \ion{O}{vi} absorption line at 21.79 \AA. 

\begin{table}
\caption{Comparison of the \ion{O}{vii} forbidden line from three different observations.}
\label{tab:como7}
\begin{tabular} {|lrrr|}\hline
Instrument & v             & $\sigma_{\rm v}$ & flux                 \\
           & (km s$^{-1}$) & (km s$^{-1}$) & (ph m$^{-2}$ s$^{-1}$) \\\hline
HETGS      & $-70\pm100$   & $<$310        & 0.8$\pm$0.2            \\
LETGS      & 0$\pm$50      & $<$330        & 0.8$\pm$0.3            \\
RGS        & $-110\pm80$   & $<$490        & 0.9$\pm$0.1            \\\hline
\end{tabular}\\
\end{table}
\par
Another emission feature for oxygen is the radiative recombination continuum (RRC) of \ion{O}{vii} between 16.9 and 17 \AA~(see Fig.~\ref{fig:uta}). As there is severe blending with iron absorption lines, we approximated the RRC simply with a delta function. We find a 2 $\sigma$ detection for this RRC and the narrowness of the feature (unresolved by the instrument) indicates a low temperature. The flux is 0.06 $\pm$ 0.03  ph m$^{-2}$ s$^{-1}$ which is consistent with the 0.06 $\pm$ 0.10 ph m$^{-2}$ s$^{-1}$ value as determined from the LETGS spectrum (Kaastra et al. \cite{kaastra02}).
\par
The presence of inner-shell absorption lines of oxygen together with the higher ionized \ion{O}{vii} and \ion{O}{viii} absorption lines allow for an accurate ionization determination, independent of elemental abundances. The ionization parameter determined for oxygen spans three orders of magnitude from log $\xi = -1.8$ for \ion{O}{iii} to log $\xi = +1.8$ for \ion{O}{viii}.
\par
The deepest absorption line in the spectrum is the \ion{O}{viii} Ly$\alpha$ line. For \ion{O}{viii} we also detect the Ly$\beta$, Ly$\gamma$ and Ly$\delta$ lines and some hints for the Ly$\epsilon$ line in absorption (see Fig.~\ref{tab:o1fit}). The multitude and depth of these absorption lines allow thus for an accurate column density and outflow velocity determination (see Table~\ref{tab:slabfit}). These higher order lines are important for the detection of variability in the warm absorber, as they have an optical depth close to unity. However, we did not detect significant variability in the measured column densities as compared to the previous LETGS observation, which has rather large error bars. Also within this observation no variability in the warm absorber was detected.
\par
Fig.~\ref{fig:oreg} shows that our model overestimates the \ion{O}{vii} resonance line. A probable explanation is that the emission component of this line partly refills the absorption line due to the imperfect spectral resolution. In the LETGS data (Kaastra et al. \cite{kaastra02}) this line has a clear P Cygni profile, and also in the RGS there is excess emission at wavelengths just longer than the resonance line. Both models for the warm absorber don't include possible P Cygni profiles. The excess emission just long-ward of 22.25 \AA~is noise, as the width of the feature is narrower than the point spread function of the instrument. The Galactic absorption by \ion{O}{vii}, detected with 1.5 $\sigma$ significance is also indicated in Fig.~\ref{fig:oreg}.
\par
Shortward of the forbidden of \ion{O}{vii} there is a deep \ion{O}{vi} absorption line, and 0.25 \AA~longward of the forbidden line there is a deep \ion{O}{v} absorption line. Both are inner-shell absorption  lines, indicating the importance of these transitions in AGN. For \ion{O}{v} two absorption lines are detected. However, the line at 22.33 \AA~is redshifted by about 700 km s$^{-1}$ relative to the weaker line at 19.92 \AA, and the other oxygen absorption lines. To check the possible uncertainty on the wavelength for the strongest \ion{O}{v} line, the \ion{O}{v} wavelength and oscillator strength were also calculated using the Cowan code (\cite{cowan}) (Raassen, private communication). This gives a rest wavelength of 22.38 \AA, closer to the observed rest wavelength of 22.39 \AA, and indicates that uncertainties in the rest wavelength can be of order 0.05 \AA; i.e. equivalent to the instrumental FWHM.

\subsection{Absorption by iron}
\begin{figure}
\resizebox{\hsize}{!}{\includegraphics[angle=-90]{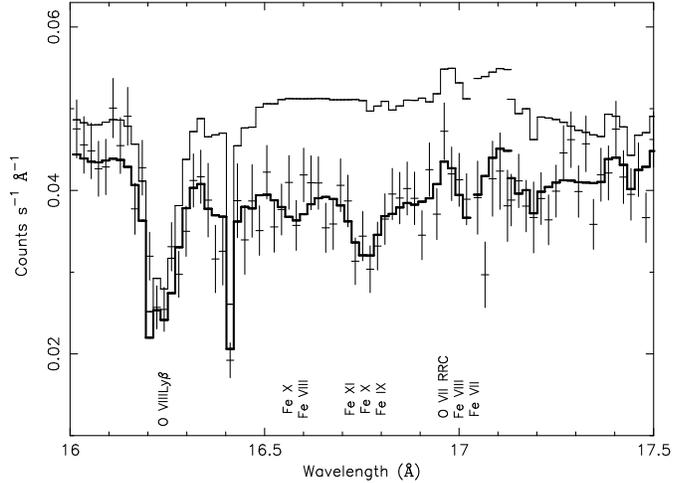}}
\caption{Detail of the first order data showing the UTA between 16 and 17.5 \AA. The thick line is the best fit through the data, the thin line indicates the same model if we set the iron abundance for the lowest ionization component to 0. A dead pixel occurs at 16.4 \AA. The gaps in the model at 17.05 and 17.15 \AA~are due to missing data. Only the deepest features are labeled.}
\label{fig:uta}
\end{figure} 
Iron is the best indicator for the ionization structure of the warm absorber, as it has absorption lines in the RGS band from neutral \ion{Fe}{i} to the highly ionized \ion{Fe}{xxiv}.  For NGC~5548 we detect absorption from \ion{Fe}{vi} to \ion{Fe}{xxiv} (see Table~\ref{tab:slabfit}). The lowest ionization component is represented by the UTA (Sako et al. \cite{sako}) formed from inner-shell transitions of \ion{Fe}{vi} through \ion{Fe}{xi}. These ions represent the lower ionized iron ions. Fig.~\ref{fig:uta} gives a detailed view of the spectrum between 16 and 17.5 \AA, where the UTA is located. Note that although there are some individual absorption features visible, the dominant effect of the UTA is the depression of the continuum due to an unresolved blend of transitions with small cross-sections. The UTA clearly shows the importance of these inner-shell transitions in determining the continuum correctly.
\par
The moderately ionized component is represented by \ion{Fe}{xiii} through \ion{Fe}{xix}, where we only find an upper limit for \ion{Fe}{xii} and \ion{Fe}{xiv}. The highest ionized component is represented by absorption by \ion{Fe}{xix} through \ion{Fe}{xxiv}. For \ion{Fe}{xix} through \ion{Fe}{xxii} we detect a multitude of absorption lines between 8 (lowest wavelength that we included in the fit) and 15 \AA. For \ion{Fe}{xxiii} and \ion{Fe}{xxiv} only two absorption lines are detected, resulting in uncertain velocity determinations.
\par
It is thus clear that all three ionization components as fitted with {\it xabs} are amply confirmed by iron absorption. The question arises whether a continuous ionization structure would be a better fit to the data. From \ion{Fe}{vi} through \ion{Fe}{xxiv} we detect all iron ions with the exception of \ion{Fe}{vii}, \ion{Fe}{xii} and \ion{Fe}{xiv}. Between \ion{Fe}{xii} and \ion{Fe}{xv} all column densities are low. Forcing \ion{Fe}{xii} to \ion{Fe}{xvi} to have a similar column densities as \ion{Fe}{xi} or \ion{Fe}{xvii} worsens the fit, increasing ${\chi_\nu}^2$ by 0.24. This indicates that for log $\xi$ between 1 and 1.4 ($\xi$ between 10 and 25) (in 10$^{-9}$ W m) the column densities are lower than for the other ionization states. However, no clear depression in column densities is seen for the transition between the middle and highly ionized component. A more continuous ionization structure than presented by the {\it xabs} model for the warm absorber is not excluded by the data.

\subsection{Absorption by carbon and nitrogen}
\begin{figure}
\resizebox{\hsize}{!}{\includegraphics[angle=-90]{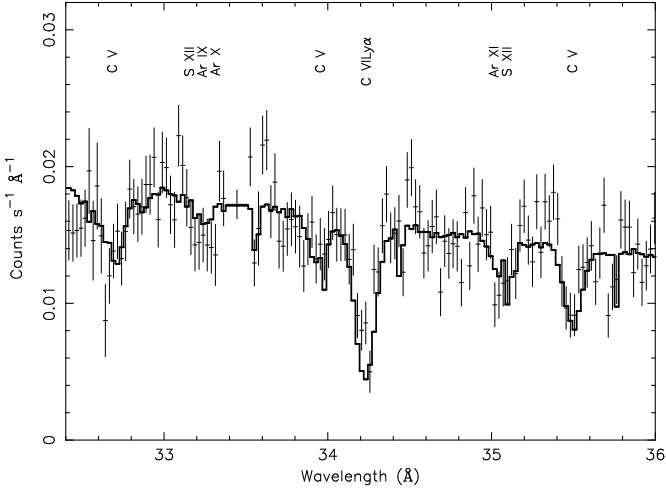}}
\caption{Detail of first order RGS 1 and 2 data, indicating the \ion{C}{vi} Ly$\alpha$ absorption line as well as some \ion{C}{v}, \ion{S}{xii}, \ion{Ar}{ix}, \ion{Ar}{x} and \ion{Ar}{xi} absorption lines. A dead pixel is seen at 34 and 35.8 \AA.}
\label{fig:creg}
\end{figure}
\begin{figure}
\resizebox{\hsize}{!}{\includegraphics[angle=-90]{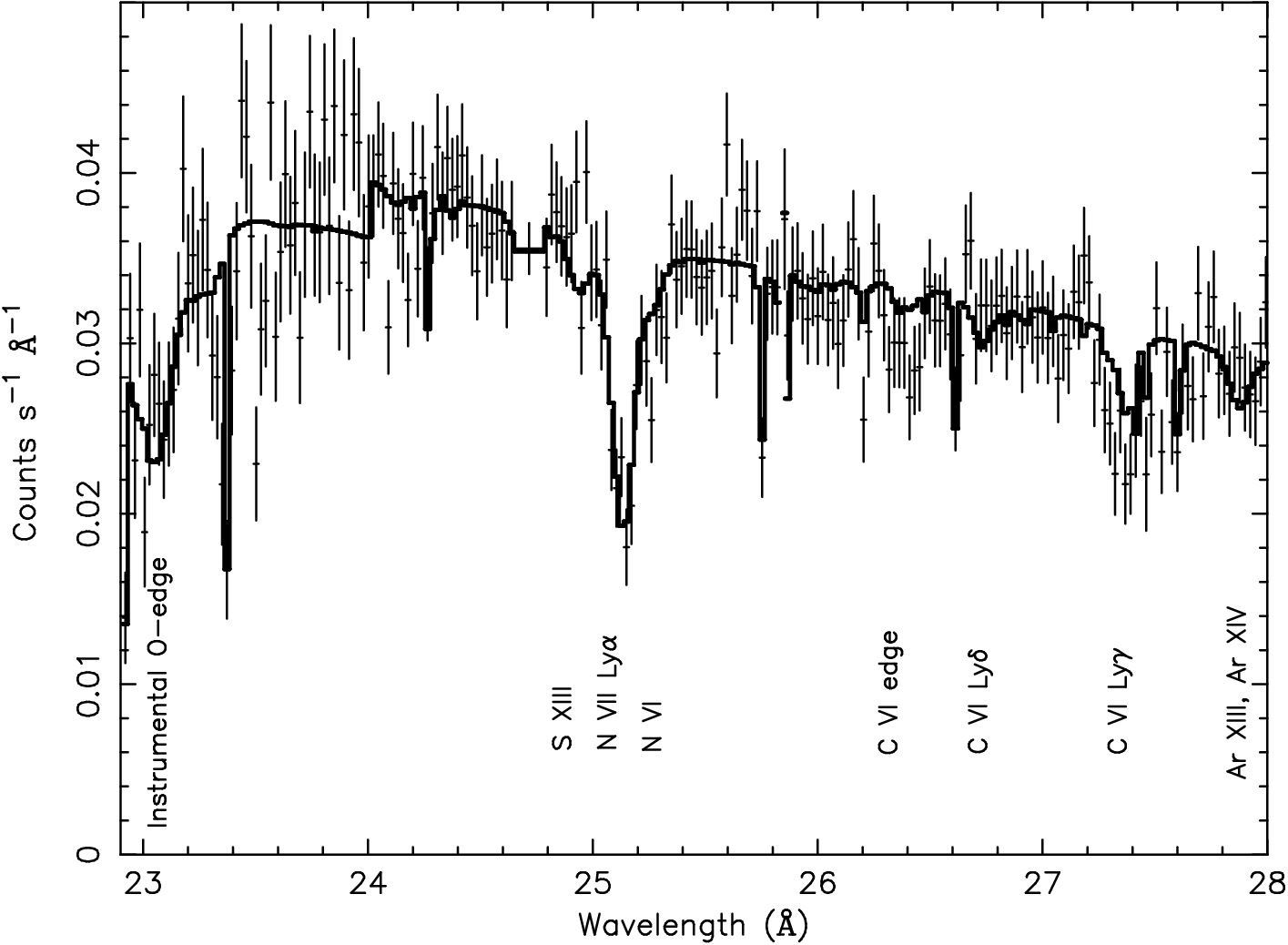}}
\caption{Detail of first order RGS 1 and 2 data, indicating the \ion{N}{vii} Ly$\alpha$ absorption line as well as the \ion{C}{vi} Ly$\gamma$ and Ly$\delta$ lines. Absorption of \ion{S}{xiii}, \ion{Ar}{xiii}, and \ion{Ar}{xiv} are labeled. Dead pixels are seen at 23.4 \AA, 24.3 \AA, 25.7 \AA, 26.2 \AA, 26.6 \AA, and 27.6 \AA.}
\label{fig:nreg}
\end{figure}
\begin{figure}
\resizebox{\hsize}{!}{\includegraphics[angle=-90]{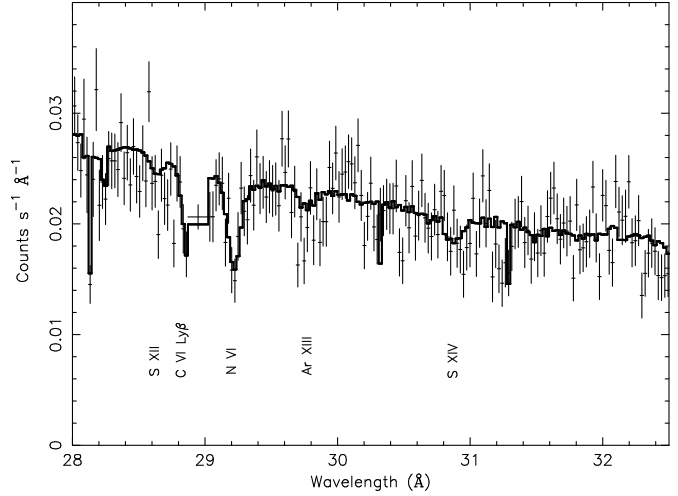}}
\caption{Detail of first order RGS 1 and 2 data, indicating an \ion{N}{vi} absorption line as well as the \ion{C}{vi} Ly$\beta$ line. Further absorption by \ion{S}{xii}, \ion{S}{xiv} and \ion{Ar}{xiii} are labeled. Dead pixels are seen at 28.1 \AA~and 30.3 \AA. A CCD gap related feature is seen at 28.9 \AA.}
\label{fig:n6reg}
\end{figure}

For carbon we detect a strong \ion{C}{vi} Ly$\alpha$ line (see Fig.~\ref{fig:creg}), the Ly$\beta$ line (see Fig.~\ref{fig:n6reg}), Ly$\gamma$ and hints of Ly$\delta$ lines (see Fig.~\ref{fig:nreg}). These higher order lines are important in the correct determination of the column density as well as any possible time variability, as their optical depth is around unity. \ion{C}{vi} and \ion{N}{vii} are produced mainly in the middle ionization component, while \ion{C}{v} and \ion{N}{vi} are produced in the middle and lowest ionization component. Finally, \ion{C}{iv} and \ion{N}{v} are produced only in the lowest ionization component (see Table~\ref{tab:slabfit} for an indication of the ionization state). The \ion{C}{vi} Ly$\alpha$ line is rather deep and at a relatively long rest wavelength of 33.736 \AA, resulting in a very good velocity shift determination. The velocity component found from the X-rays is consistent with component 3 from the UV, which has an outflow velocity of $-540$ km s$^{-1}$ instead of $-510$ km s$^{-1}$. There is a hint in our data that the line has substructure, consistent with the earlier LETGS results.

\subsection{Relativistically broadened emission lines}
\begin{table}
\caption{Values for the relativistically broadened emission lines in NGC~5548, MCG$-$6-30-15 and Mrk~766 (Sako et al. \cite{sako}). LETGS measures are taken from Kaastra et al. (\cite{kaastra02}).}
\label{tab:relfit}
\begin{tabular} {|l|rrrr|}\hline
    &   RGS      & LETGS       & MCG$-$6-30-15 & Mrk~766 \\
i$^{a}$ & 46$\pm$35 & 46($-$3,+8) & 38.5$\pm$0.4  & 34.3$\pm$0.74 \\
q$^{b}$ & 3.9$\pm$5  & 3.9$\pm$0.6 & 4.49$\pm$0.15 & 3.66$\pm$0.22 \\
r$_{\rm in}$$^{c}$ & 100$\pm$30  &$<$2.6       &3.21$\pm$1.2   & $<$2.3\\
r$_{\rm out}$$^{c}$& 400$\pm$90  &$>$10        &100($-$48,+95)&81($-$20,+73)\\
\ion{C}{vi}$^{d}$& $<$0.8& $<$0.7   &2.0$\pm$0.2   & 2.3$\pm$0.2\\
\ion{N}{vii}$^{d}$&$<$0.03& 1.1$\pm$0.4     &4.4$\pm$0.2   & 5.8$\pm$0.3\\
\ion{O}{viii}$^{d}$& 0.2$\pm$0.1& 0.6$\pm$0.2  &13.1$\pm$0.6  & 11.4$\pm$0.7\\\hline
\end{tabular}\\
$^{a}$ The inclination derived for the AGN from the relativistic lines, in degrees.\\
$^{b}$ The emissivity slope.\\
$^{c}$ In GM/c$^{2}$.\\
$^{d}$ EW measured in \AA.
\end{table}
After finding a model for the warm absorber that is as complete as possible, we added relativistically broadened emission lines of \ion{O}{viii}, \ion{N}{vii}, and \ion{C}{vi} Ly$\alpha$ to the model. Such lines were observed in MCG$-$6-30-15 and Mrk~766 (Branduardi-Raymont et al. \cite{brand}, Sako et al. \cite{sako02}). 
\par
\begin{figure}
\resizebox{\hsize}{!}{\includegraphics[angle=-90]{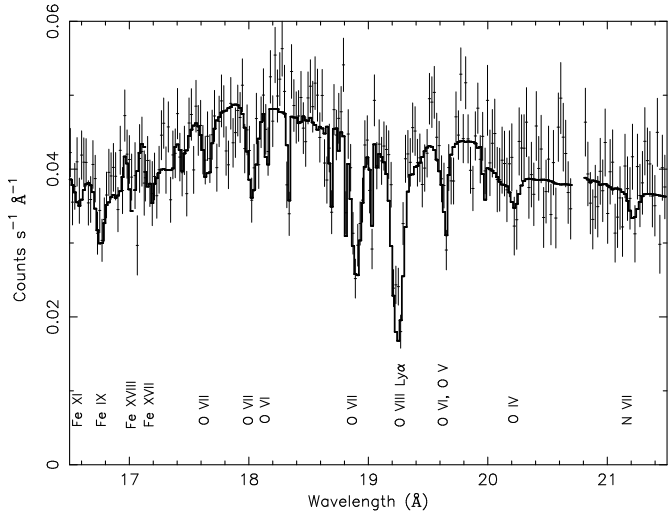}}
\caption{Best {\it xabs} fit to the first order data, detailing the wavelength band around the position of the relativistic \ion{O}{viii} Ly$\alpha$ emission line. Note that in this plot no relativistic line is fitted. Dead pixels are seen at 17.4 \AA, 18.3 \AA, 18.7 \AA, 18.8 \AA, 19 \AA, and 19.95 \AA. A CCD gap related feature can be seen between 20.7 and 20.8 \AA.}
\label{tab:rel}
\end{figure}

 
In NGC~5548 Kaastra et al. (\cite{kaastra02}) found evidence for weak broadened emission lines of \ion{O}{viii} Ly$\alpha$ and \ion{N}{vii} Ly$\alpha$. However, no indication for a relativistically broadened \ion{C}{vi} Ly$\alpha$ line was detected. We applied the best fit parameters found from the LETGS model to the RGS data but found no significant improvement for any of the relativistically broadened emission lines. Leaving the parameters for the Laor line profile (Laor \cite{laor}) free, we find that a broad Gaussian would better describe the line/excess. However, the improvement in fit is still small for all three emission lines. The results are given in Table~\ref{tab:relfit}, together with the best fit values for the LETGS data of NGC~5548 (Kaastra et al. \cite{kaastra02}), and the RGS data of MCG$-$6-30-15 and Mrk~766 (Branduardi-Raymont et al. \cite{brand}, Sako et al. \cite{sako02}). Note, that the values found for the EW between our RGS results and the LETGS results are consistent within 3 $\sigma$, except for the inner radius. These relativistic lines should be time variable, considering that they are formed in the accretion disk close to the black hole. Possibly, time variability explains the differences between the previous LETGS results and ours. Given their weakness, in the further analysis and in all the plots we did not include these broadened emission lines. Fig.~\ref{tab:rel} gives the wavelength band where the relativistic \ion{O}{viii} Ly$\alpha$ line should be situated, and a small excess is seen between 18 and 18.4 \AA. 

\section{Discussion}

\subsection{Outflow velocity versus ionization}

\begin{figure}
\resizebox{\hsize}{!}{\includegraphics[angle=-90]{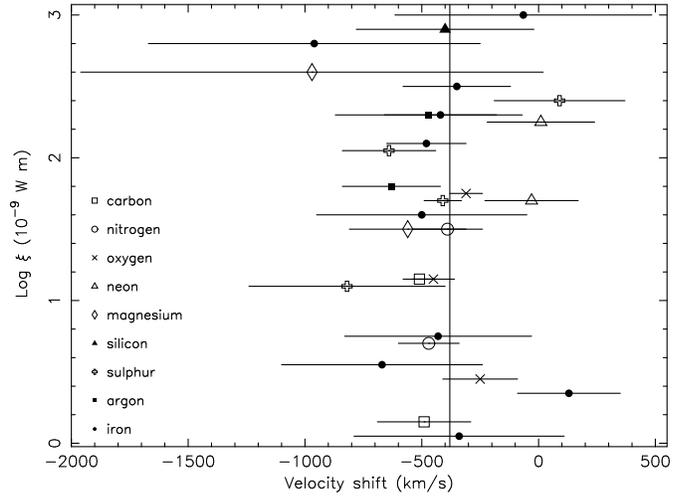}}
\caption{The velocity shift as given in Table~\ref{tab:slabfit} versus the the ionization parameter. The vertical line represents the average outflow velocity.}
\label{fig:xizv}
\end{figure}

No evidence is found for a correlation between the outflow velocity and the ionization parameter (see Fig.~\ref{fig:xizv}). Rather the outflow velocity, for first order, is consistent with $-$380 km\,s$^{-1}$ and $-$350 km\,s$^{-1}$ for the {\it slab} and {\it xabs} model respectively over the three orders of magnitude sampled in ionization scale. This is consistent with the UV data (Crenshaw \& Kraemer \cite{crenshaw}), and the results for the \ion{C}{vi} Ly$\alpha$ line in the LETGS data (Kaastra et al. \cite{kaastra02}). Kaastra et al. (\cite{kaastra02}) found a trend of decreasing outflow velocity for higher ionized iron and oxygen ions, although all outflow velocities were consistent with an outflow velocity of $-$340 km s$^{-1}$. This result is not reproduced here. The outflow velocities observed in both data sets are consistent with the five UV velocity components, suggesting a similar origin of both absorbers.

\subsection{Hydrogen column density versus ionization}
\begin{figure}
\resizebox{\hsize}{!}{\includegraphics[angle=-90]{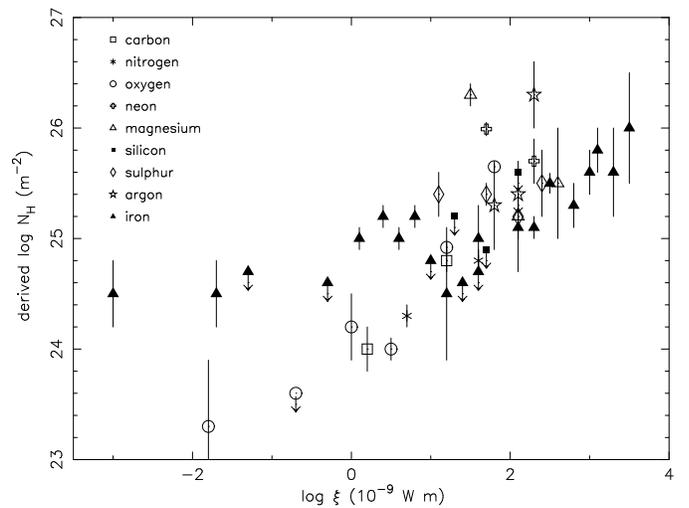}}
\caption{The derived absorbing column density per ion as derived from the column densities given in Table~\ref{tab:slabfit}. There is an increase of the derived column density with higher ionization.} 
\label{fig:colion} 
\end{figure}

The derived absorbing column density per ion, as determined from the {\it slab} modeling, versus the ionization parameter is plotted in Fig.~\ref{fig:colion}. In the Figure we use the ionization parameter corresponding to the state with maximum column density. In reality, however, the ion is produced over a range of ionization states. Fig.~\ref{fig:colion} thus assumes a continuous ionization distribution, and is dependent on the spectral energy distribution (SED) in our models with XSTAR. We tested a model with and without a UV bump, and found a maximum displacement in log $\xi$ of 0.3 for very highly ionized iron, and a maximum change in the derived absorbing column density of 0.1 on the log scale used. However, as most ions at a particular ionization state have a similar shift, the derived abundances and the derived absorbing column density are rather insensitive to the input SED. 
\par
In general, the derived absorbing column density increases for higher ionization parameters. The correlation coefficient, for all data points, excluding upper limits, is 0.71, and the related significance is 0.999997; for iron these values are 0.82 and 0.9982, respectively. The derived absorbing column density from the three component {\it xabs} model is very similar to the absorbing column density as derived from the {\it slab} model, and a similar increase for higher ionization is observed.
\par
In general the abundances found are consistent with solar. An exception is iron, for which both models have lowly ionized iron overabundant compared to lowly ionized oxygen, while a reverse trend occurs for highly ionized iron and oxygen. As carbon and nitrogen are consistent with oxygen, the effect is probably due to uncertainties for the lowly ionized iron ions.  A similar overabundance for iron in the lowest ionization state was noted by Blustin et al. ({\cite{blustin}) for NGC~3783. In the LETGS observation of NGC~5548 (Kaastra et al. \cite{kaastra02}) found a similar abundance pattern for iron, throughout the ionization range. The explanations listed by Kaastra et al. (\cite{kaastra02}) for this discrepancy have been tested using this data set. First, a calibration effect is unlikely as RGS and LETGS give consistent results. Secondly, we determined the column density for lowly ionized oxygen ions from our own X-ray spectrum; and thus we do not need to rely on non-simultaneous UV measurements. Spectral variability is thus not a possible explanation for the apparent overabundance of lowly ionized iron. Thirdly, we tested the sensitivity of the model to different values for the line broadening, ensuring that saturation due to narrow line widths cannot be responsible for the overabundance of lowly ionized iron. Reducing the width of the lowly ionized absorption lines, has the effect that iron saturates first, worsening the overabundance already found. The maximum width of the lowly ionized lines is set by the width of the inner-shell oxygen lines, which are unresolved.
\par 
Possible explanations for the iron overabundance are an uncertainty in the wavelength of the M-shell iron absorption lines; or if the optical depth, $\tau\gg$ 1, an uncertainty in the radiative transfer models used. Another possibility is the omission of certain processes like re-emission and Bowen fluorescence in the radiative transfer models used. Finally, the iron abundance could be non-solar, although the physical process causing such an overabundance of iron in Seyfert galaxies is unknown. Non-solar abundances have been detected in IRAS~13349+2438 (Sako et al. \cite{sako}) and NGC~1068 (Kinkhabwala et al. \cite{kink}).}

\subsection{Outflow geometry}

The geometry and physics of the warm absorber as observed in some AGN is poorly understood. One of the outstanding questions is whether this outflow is spherical, in which case the ionization of the gas is mainly dependent on the distance. Alternatively, the outflow could be in localized streams formed due to instabilities in the accretion disk. The ionization parameter is then mainly dependent on the density. As distance and density cause a degeneracy in the determination of the ionization parameter, the location of the outflowing X-ray and UV warm absorber is uncertain. Constraining the geometry of this outflow will help solve these outstanding problems.
\par
From mass conservation ($\dot{M}_{\rm loss}$ = 4 $\pi r^2 n m_{\rm p} v \Omega$), the ionization parameter ($\xi$ = $L/n r^2$), the outflow velocity ($v$), and the bolometric luminosity ($L$), one can determine the mass loss rate, $\dot{M}_{\rm loss}$, as a function of the opening angle ($\Omega$) of the outflowing wind. This is summarized in eq. (\ref{eq:1}), where $m_{\rm p}$ is the proton mass. Assuming that the mass loss rate of the wind cannot be higher than the mass accretion rate onto a black hole and that the system is stationary, one can equate both and obtain an upper limit for the opening angle. Eq.~(\ref{eq:2}) gives the mass accretion rate, $\dot{M}_{\rm acc}$, with a mass conversion efficiency of $\eta$.
\begin{equation}
\label{eq:1}
\dot{M}_{\rm loss} = \frac{4 \pi L m_{\rm p} v \Omega}{\xi}
\end{equation}
\begin{equation}
\label{eq:2}
\dot{M}_{\rm acc} = \frac{L}{c^2\eta}
\end{equation}
In the case of a Schwarzschild black hole, as considered here, $\eta$ = 0.057. For our average outflow velocity of 380 km s$^{-1}$ the upper limits of the opening angle vary between 3.8 $\times$ 10$^{-7}$ sr for log $\xi$ = $-$1.8 and 0.024 sr for log $\xi$ = 3, with log $\xi$ in 10$^{-9}$ W m. For an average ionization state of log $\xi$ = 1.7 in 10$^{-9}$ W m, we find an upper limit to the opening angle of 0.0019 sr. The upper limits to the opening angle indicate that the outflow is mostly in a narrow stream, where the densest part of the stream is the narrowest and lowest ionized. For higher outflow velocities or a higher mass conversion efficiency, i.e. in the case of a Kerr black hole, the stream is even narrower.
\par
From Fig.~\ref{fig:colion} we determine that the derived absorbing column density scales with the ionization parameter, $N_{\rm H}$ $\sim$ $\xi^{\alpha}$. $\alpha$ ranges from 0.25 to 0.5 depending on whether one uses the $N_{\rm H}$ obtained for the iron or oxygen ions at the lower ionization states. We define $N_{\rm H} \equiv n d$, where $n$ is the density and $d$ is the thickness of the outflow in the line of sight. Equating $N_{\rm H}$ to $\xi^{\alpha}$, where $\xi$ = $L/n r^2$, one can determine a relation between the density, thickness and distance. If one further assumes that for a fixed distance ($r$) from the ionizing source, there is a range in densities, resulting in a range in ionization states, one finds that the density is related to the thickness $n \sim d^{-5/4}$ or $n \sim d^{-3/2}$, respectively. From both simple models we conclude that the outflow occurs in streams with a very small opening angle for the dense and lowly ionized gas, and with a larger opening angle for the less dense and higher ionized gas.

\subsection{Comparison of UV and X-ray detection of \ion{O}{vi}}
\ion{O}{vi} has absorption lines in both the UV ($\lambda\lambda$1032, 1038 \AA) and X-ray band (22.01 \AA, 19.34 \AA~and 21.79 \AA). Our X-ray column density of \ion{O}{vi} of 10$^{20.5}$ m$^{-2}$ is nearly an order of magnitude larger than the UV column density (10$^{19.62}$ m$^{-2}$) of Brotherton et al. (\cite{brotherton}). However, the outflow velocities in the UV and the X-rays are comparable. The difference in derived absorbing column density indicates that the UV data are severely saturated, leading to an underestimate in the column density. Arav et al. (\cite{arav}) compare the column density of \ion{O}{vi} as observed in this data set and an earlier FUSE UV measurement, and conclude that the column densities could be equal if velocity dependent covering is considered. However, simultaneous observations are necessary to exclude the possibility of column density variability (see Arav et al. \cite{arav} for more details). Generally, in the X-rays, we find significantly more low ionization gas than previously deduced from UV data. The discovery of \ion{O}{iii} to \ion{O}{v} absorption lines, in addition to \ion{O}{vi}, gives us for the first time an estimate for the column densities of these ions, which are inaccessible in the UV.

\section{Summary}
We have presented here the highest signal to noise high resolution X-ray spectrum of NGC~5548 obtained yet. The spectrum shows a rich structure in narrow and broad spectral features. We detected a very weak \ion{O}{vii} RRC, consistent with a low temperature. Inner-shell oxygen lines, together with the higher order absorption lines from \ion{O}{viii} Ly$\alpha$ and \ion{C}{vi} Ly$\alpha$ have an optical depth near unity, important in the detection of spectral variability of the warm absorber, although no spectral variability was detected in this data set or from comparison with the earlier LETGS spectrum. Uncertainties in rest wavelengths for the inner-shell ions complicate the study of these absorption lines, certainly for the lowly ionized ions. Possibly this or uncertainties and the omission of certain processes in the radiation transfer models lead to an deviant iron to oxygen abundance ratio.
\par
The detected warm absorber spans three orders of magnitude in ionization, from ions as lowly ionized as those measured in the UV band, to nearly fully ionized iron. Only a few percent of the X-ray warm absorber column density is lowly ionized, represented by oxygen and iron inner-shell transition lines. The outflow velocities determined from the X-rays are all consistent with those measured in higher resolution UV spectra, and no correlation between the ionization state and velocity was detected. The similarity in outflow velocity and the detection of \ion{O}{vi} give strong evidence that the UV and X-ray warm absorber are different manifestations of the same outflowing wind phenomenon. However, we detect an order of magnitude more \ion{O}{vi} in the X-rays than previous UV measurements. In general, we detect substantially more low ionized gas, than is deduced to be present from UV observations. Probably the UV band underestimates the true column densities due to saturation and velocity dependent covering factor. From simple arguments, we found that the outflow detected is consistent with a narrow stream, with smallest opening angles for the densest and lowest ionized gas.

\section*{ACKNOWLEDGMENTS}
 
This work is based on observations obtained with XMM-{\it Newton}, an ESA science mission with instruments and contributions directly funded by ESA Member States and the USA (NASA). SRON National Institute for Space Research is supported financially by NWO, the Netherlands Organization for Scientific Research. RE acknowledges support from the NASA XMM-{\it Newton} grant NAG5-10032. We thank Ehud Behar (Columbia University), Ton Raassen (SRON) for helpful discussions on the uncertainties of inner-shell absorption line wavelengths.

\end{document}